\documentclass[12pt]{article}
\usepackage{amsmath,amssymb,amsfonts}
\usepackage[dvips]{graphicx}
\usepackage{epsfig}

\usepackage[vcentermath]{youngtab}

\def\beq{\begin{equation}}
\def\eeq{\end{equation}}
\def\bea{\begin{eqnarray}}
\def\eea{\end{eqnarray}}

\makeatletter \@addtoreset{equation}{section} \makeatother

\begin{document}

\renewcommand{\thefootnote}{\alph{footnote}}

\begin{titlepage}

\begin{center}
\hfill {\tt SNUTP10-004}\\
\hfill {\tt arXiv:1003.4343}

\vspace{2cm}

{\Large\bf Probing AdS$_4$/CFT$_3$ proposals beyond chiral rings}

\vspace{2cm}

\renewcommand{\thefootnote}{\alph{footnote}}

{\large Seok Kim$^1$ and Jaemo Park$^{2}$}

\vspace{1cm}

\textit{$^1$Department of Physics and Astronomy \& Center for
Theoretical Physics,\\
Seoul National University, Seoul 151-747, Korea.}\\

\vspace{0.2cm}

\textit{$^2$Department of Physics \& Postech Center  for Theoretical
Physics (PCTP),\\
POSTECH, Pohang 790-784, Korea}\\

\vspace{0.7cm}

E-mails: {\tt skim@phya.snu.ac.kr, jaemo@postech.ac.kr}

\end{center}

\vspace{1cm}

\begin{abstract}

We calculate the superconformal Witten index for the
Chern-Simons-matter theory which was proposed to describe multiple
M2-branes on $\mathbb{C}^2\times\mathbb{C}^2/\mathbb{Z}_k$. We
consider a variant of this model, which exhibits
explicit $\mathcal{N}\!=\!3$ supersymmetry and has the advantage
of not having an exotic branch of the moduli space. At $k=1$, we
compare the index with that from the proposed gravity dual and
find a disagreement.

\end{abstract}

\end{titlepage}

\renewcommand{\thefootnote}{\arabic{footnote}}

\setcounter{footnote}{0}

\tableofcontents

\section{Introduction}
Recently, we have seen the tremendous progress in understanding
$AdS_4/CFT_3$ correspondence. The gravitational part involving
$AdS_4$ was relatively well-known. The harder problem was to
understand $CFT_3$ corresponding to M2 branes probing a part of
Calabi-Yau 4-fold. The crucial observation was made by Schwarz
\cite{sch} that these $CFT_3$s can be described by Chern-Simons
gauge theory with matter without the usual Yang-Mills kinetic
terms. In retrospect, the ${\cal N}=8$ theory of
Bagger-Lambert-Gustavsson \cite{BL1, BL2, BL3,gus1, gus2} is the
avatar of this idea, which is equivalent to $SU(2)\times SU(2)$
gauge theory \cite{rams}. More general classes of ${\cal N}=4$
Chern-Simons theories describing M2 brane probing suitable
non-compact Calabi-Yau manifold were constructed by \cite{HLLLP}
following the method of \cite{GaiottoWitten}. It turns out that
the special case of the construction of \cite{HLLLP} describes M2
branes probing $C^4/Z_k$, whose AdS/CFT aspects were worked out in
detail in \cite{ABJM}, which is called ABJM theory with enhanced
${\cal N}=6$ supersymmetry \cite{HLLLP2, Schwarz2}.

The analysis of ${\cal N}=2$ theories were initiated in
\cite{Martelli,Ami}. This is an interesting arena to work out
since ${\cal N}=2$ theories correspond to M2 branes probing
general Calabi-Yau four-folds. The curious fact of ${\cal N}=2$
theory is that there could be multiple Chern-Simons gauge theories
having the same moduli space. The typical example is the so-called
the dual ABJM theory which is $U(N) \times U(N)$ ${\cal N}=2$
theory with bifundamentals $A \, (N, \bar{N})\,\,$,  $B \,
(\bar{N}, N)$ and with additional adjoint fields $\phi_1, \phi_2$
of the first gauge group factor. With the superpotential $W={\rm
Tr} AB[\phi_1, \phi_2]$ and with the Chern-Simons level $(k, -k)$,
one can show that the moduli space contains the symmetric product
of $C^2/Z_k \times C^2$.\footnote{Another model with moduli space
of $C^2/Z_k \times C^2$ is constructed by using Chern-Simons
matter theories with chiral flavors \cite{Benini}. Also in
\cite{nkim} dual ABJM model is obtained via orbifold procedure of
Nambu bracket.} For $k=1$ the moduli space has the symmetric
product of $C^4$. One might suspect that the ABJM and the dual
ABJM theory could have the identical gravitational dual for $k=1$.
On the other hand, it is observed that there are at least 19
models of ${\cal N}=2$ theory which has the symmetric product of
$C^4$ as the moduli space \cite{HeSparks}. Thus the important
question is if all of these models are describing M2 branes
probing $C^4$. Since all of these theories involve the
Chern-Simons level $k=0,1$, it's rather difficult to analyze these
theories perturbatively. There'a a possibility that some of these
theories might be related via Seiberg-duality.

In order to answer this question, we should work out more than the
chiral ring or moduli space of the underlying theories. Recently
there were several attempts along this direction. One is the
compuatation of the superconformal Witten index \cite{Bhatta, Kim:2009wb},
which provides the detailed information of BPS states (or local operators)
with lower supersymmetry. Another is the computation of the partition function and the
Wilson loop using the localization technique \cite{Kapustin,
Drukker}. Especially in \cite{Marino}, 1/2 BPS Wilson loop is
evaluated exactly in ABJM theory. The goal of this letter is to
compare the indices of two theories which have the same moduli
space. The theories we are interested in are ABJM and ${\cal N}=3$
variant of the dual ABJM theory which we introduce in section 2. The moduli space of the ${\cal N}=3$
variant, with Chern-Simons level $(k, -k)$, is given by the symmetric product of $C^2/Z_k \times C^2$.
Furthermore it does not have additional exotic branches, in
contrast to ${\cal N}=2$ dual ABJM model \cite{RodriguezGomez:2009ae}.
Besides, we believe that the ${\cal N}=3$ theory is
better suited for the index computation for technical reasons. One
can trust the various assumptions made in the index computation
while for ${\cal N}=2$ theories, this does not seem to be the case, on which we will comment later.
In particular, one can carry out the index computation for
Chern-Simons level $(k,-k)=(1,-1)$ where the underlying moduli
space is the symmetric product of $C^4$. One can see that the
index computation of ${\cal N}=3$ theory leads to the different
result than ABJM case with $(k,-k)=(1,-1)$. Thus one can show
explicitly that having the same moduli space does not necessarily
give the same M2 brane CFT. Given this result, one might wonder
what is the corresponding gravitational dual of the ${\cal N}=3$
theory or whether this ${\cal N}=3$ theory has the gravitational dual at
all. We will also explain a deconfinement behavior of the index,
which seems to be in tension with the M2 brane interpretation.

One may ask similar questions for various ${\cal N}=2$ theories
having the symmetric product of $C^4$ as the moduli space, though
we cannot show any computation for these theories. One interesting
example related to this question is again found in ${\cal N}=6$
theory. If we consider ${\cal N}=6$ $U(N+l)_k \times U(N)_{-k}$
theory with $l\leq k$, they have the same moduli space
$Sym^N(C^4/Z_k)$ but they have different gravitational duals
distinguished by the different discrete flux data in the
gravitational side \cite{ABJ}. We hope to understand if various
${\cal N}=2$ theories are related in a similar way or another, but
this is beyond the scope of the current work. Finanlly, the
current work has an interesting implication for the relation
between the crystal models and the Chern-Simons matter theories in
(2+1)-dimensions \cite{Sangmin1,Sangmin2,Sangmin3}. There are
various proposals for extracting the gauge theory from the crystal
models\cite{Sangmin3, HananyVegh}. Especially in
\cite{HananyVegh}, it is suggested that ABJM and dual ABJM model
are two theories read off from $C^4$ crystal model.  Given our
result, this proposal should be modified. And it is an interesting
topic to figure out the precise gauge theory associated with a
given crytal model, given a successful construction of various
${\cal N}=2$ theories where the corresponding crystal models
\cite{FrancoPark} play a crucial role.

\section{Dual ABJM and its variant}

A (2+1)d ${\cal N}=2$ Chern-Simons(CS) theory with bifundamental
and adjoint matter is given, in ${\cal N}=2$ superspace notation,
by the following Lagrangian \cite{Ami}
\begin{equation}
{\rm Tr}\left( -\int
d^4\theta \sum_{X_{ab}} X_{ab}^\dagger e^{-V_a}
  X_{ab} e^{V_b}  -
\sum_a \frac{k_a}{2\pi}\int_0^1 dt  V_a \bar D^\alpha (e^{t
V_a}D_\alpha e^{-t V_a})
 +  \int d^2\theta  W(X_{ab}) + {\rm c.c.}\right) ~,
\end{equation}
where $V_a$ are vector supermultiplets and $X_{ab}$ denote chiral
supermultiplets transforming in the fundamental representation of
gauge group $a$ and the anti-fundamental representation of gauge
group $b$. For $a=b$, this corresponds to adjoint matter for gauge
group $a$. We take $\sum k_a=0$. This is a necessary condition for
the moduli space to be four complex dimensional.
% since all the cases we consider will satisfy this condition.
Recall that in 2+1 dimensions a vector superfield has the
expansion
\begin{equation}
V= -2 i \theta\bar \theta \sigma + 2 \theta \gamma^\mu \bar\theta
A_\mu + \cdot\cdot\cdot + \theta^2\bar\theta^2 D ~,
\end{equation}
where we omitted the fermionic part. Compared to (3+1)-dimensions,
there is a new scalar field $\sigma$. We can write all terms
contributing to the scalar potential in the Lagrangian
\begin{equation}
{\rm Tr} \left(-  \sum_a  \frac{2 k_a}{\pi} \sigma_a D_a + \sum_a
D_a \mu_a(X)
  -\sum_{X_{ab}} (\sigma _a
X_{ab} - X_{ab} \sigma_b)(\sigma _a X_{ab} - X_{ab}
\sigma_b)^\dagger -\sum_{X_{ab}} |\partial_{X_{ab}} W|^2\right) ~.
\end{equation}
 $\mu_a(X)$ is the moment map for the $a$-th gauge group
\begin{equation}
 \mu_a(X)= \sum_b X_{ab} X_{ab}^\dagger - \sum_c
X_{ca}^\dagger X_{ca} + [X_{aa},X_{aa}^\dagger] ~,
\end{equation}
and gives the D-term. Here we use the same terminology of (3+1)d.

By integrating out the auxiliary fields $D_a$, we see that the
bosonic potential is a sum of squares. The vacua can be found by
looking for vanishing of the scalar potential. This gives rise to
a set of matrix equations
\bea
\partial_{X_{ab}} W &=& 0\nonumber\\
 \mu_a(X) &=& \frac{2 k}{\pi} k_a \sigma_a  \nonumber\\
\sigma _a X_{ab} - X_{ab} \sigma_b &=& 0 \eea The solutions to
these equations automatically satisfy $D_a=0$ and correspond to
supersymmetric vacua. F-term constraints are exactly as in the
(3+1)d case, while D-term constraints are modified.

In this letter, we consider a special class of ${\cal N}=2$ theory
where the gauge group is given by $U(N_1) \times U(N_2)$ with two
bifundamentals $A (N_1,\bar{N}_2)$ and $B (\bar{N}_1, N_2)$ and
with two adjoints $\phi_1, \phi_2$ of $U(N_1)$. Chern-Simons level
is given by $(k, -k)$. We introduce the superpotential for this
theory to obtain ${\cal N}=3$ theory. Following \cite{Gaiotto:2007qi},
we introduce $\Phi_1, \Phi_2$ auxiliary chiral superfields of the
adjoint representation of $U(N_1), U(N_2)$ respectively. Combined
with $V_a$ of ${\cal N}=2$
 vector superfield, they form ${\cal N}=4$ vectormultiplets.
${\cal N}=3$ superpotential is obtained by starting from
\begin{equation}
\int d^2\theta (-\frac{k}{4\pi} {\rm Tr} \Phi_1^2+ {\rm
Tr}\Phi_1[\phi_1,\phi_2] +B\Phi_1 A +\frac{k}{4\pi} {\rm Tr}
\Phi_2^2 + A\Phi_2 B)
\end{equation}
and integrating out $\Phi_1, \Phi_2$ so that
\begin{equation}\label{modified-ABJM}
 W= \frac{2\pi}{k} {\rm Tr} \left(2AB[\phi_1,\phi_2]+[\phi_1,\phi_2]^2\right)\ .
\end{equation}
Note that the ${\cal N}=2$ dual ABJM is given by the
superpotential \cite{HananyVegh, FrancoPark}
\begin{equation}\label{super-original}
W=\frac{4\pi}{k} {\rm Tr} AB[\phi_1, \phi_2].
\end{equation}

Assume that $N_1=N_2\equiv N$. Let's work out the moduli space of
abelian case of ${\cal N}=3$ case. The superpotential is vanishing
identically and we have $\sigma_1=-\sigma_2\equiv \sigma$ with
\begin{equation}
\mu_1=-\mu_2=\frac{2k}{\pi}\sigma.
\end{equation}
Thus the moment map determines the value of $\sigma_a$. By
imposing the integer quantization condition  of the flux
associated with $F=dA_1+dA_2$ where $A_1, A_2$ are the gauge
fields of two U(1)s, of $\frac{1}{k}$, one fixes the  periodicity
of the conjugate variable of $A_1+A_2$, which gives the usual
discrete modding\cite{tong}
\begin{equation}
A \rightarrow e^{\frac{2\pi i}{k}} A \,\,\,\ B \rightarrow
e^{-\frac{2\pi i}{k}} B
\end{equation}
while $\phi_i$ is left invariant. Thus we have the moduli space
$C^2/Z_k \times C^2$. Note that the moduli space is identical to
the dual ABJM for the abelian case. For nonabelian case,   both
${\cal N}=3$ theory and the ${\cal N}=2$ dual ABJM model have the
symmetric product of $C^2/Z_k \times C^2$ as one branch of the
moduli space. The important question is if this is the only branch
of the moduli space. In \cite{RodriguezGomez:2009ae}, it is shown
that ${\cal N}=2$ dual ABJM model has the additional branch of the
moduli space. Below, we show that the ${\cal N}=3$ theory does
not have such additional branch so that its moduli space is
simply given by the symmetric product of $C^2/Z_k \times C^2$.

We first consider the F-term condition for $\phi_1$, $\phi_2$, $M\equiv AB$:
\begin{eqnarray}
  &&[\phi_1,\phi_2]M=M[\phi_1,\phi_2]=0\ ,\label{super-1}\\
  &&[\phi_i,M+[\phi_1,\phi_2]]=0\ \ \ \ (i=1,2)\ .\label{super-2}
\end{eqnarray}
$M$ is in the adjoint representation of the first $U(N)$ gauge
group. Compared to the case studied in
\cite{RodriguezGomez:2009ae}, we have $[\phi_1,\phi_2]$ inside the
commutator of (\ref{super-2}). Subtracting the two equations in
(\ref{super-1}), we can use $GL(N)$ which complexifies first
$U(N)$ to diagonalize both $M$ and $[\phi_1,\phi_2]$. Let the two
matrices have eigenvalues $m_I$ and $\lambda_I$, respectively,
with $I=1,2,\cdots,N$. The conditions in (\ref{super-1}) imply
$m_I\lambda_I=0$, so one of the two eigenvalues for given $I$ is
always zero. Now we consider the last condition (\ref{super-2}).
In the above diagonalizing basis, $M+[\phi_1,\phi_2]$ is also
diagonal. (\ref{super-2}) implies that $\phi_i$ can be
`generically' diagonalized in the same basis.

If there turns out to be a block of equal eigenvalues in $M+[\phi_1,\phi_2]$, (\ref{super-2}) does
not constrain off-diagonal entries of $\phi_i$ in this block. Since either eigenvalues of $m_I$ or $\lambda_I$ is zero, equal eigenvalues $m_I\!+\!\lambda_I$ of $M+[\phi_1,\phi_2]$ for
different $I$ may appear in following possibilities. Firstly, some $\lambda_I$'s can be zero
while corresponding $m_I$'s are all equal. In this block, $[\phi_1,\phi_2]=0$ from $\lambda_I=0$
and $\phi_1,\phi_2$ can clearly be diagonalized simultaneously. Secondly, some $m_I$'s can be zero
while corresponding $\lambda_I$'s are all equal. In this block, we have $M=0$ so that part of $GL(N)$ is not used, while $[\phi_1,\phi_2]$ is unconstrained from (\ref{super-1}). We use $GL(N)$
to diagonalize $\phi_1$ with eigenvalues $\alpha_I$. Inserting $M\!=\!0$ to (\ref{super-2})
with $i\!=\!1$, one obtains
\begin{equation}
  [\phi_1,[\phi_1,\phi_2]]=0\ . \label{m1}
\end{equation}
In this block, the $IJ$'th element is given by
\begin{equation}
  (\alpha_I-\alpha_J)^2(\phi_2)_{IJ}=0  \label{m2}
\end{equation}
so that the off-diagonals of $\phi_2$ is either zero or $\phi_2$
can be diagonalized if some eigenvalues of $\phi_1$ are equal.
Finally, one would worry about the blocks in which
$m_I=\lambda_J\neq 0$ with $m_J=0=\lambda_I$ (for $I\neq J$), so
that the eigenvalues of $M+[\phi_1,\phi_2]$ are still equal in
that block. For the two sub-blocks with nonzero $M$ and zero $M$,
we have diagonalized the fields and all that matter are
off-diagonal elements $(\phi_i)_{IJ}$ with $m_I=\lambda_J$,
$m_J=0=\lambda_I$ (for $I\neq J$). However, we already know that
$\lambda_J=0$ since $\phi_i$ are all diagonalized by eq.
(\ref{m1}) and (\ref{m2}), so the last case actually does not
exist.

Having diagonalized $M,\phi_1,\phi_2$ using the first $U(N)$ gauge group, one can use the second $U(N)$ to diagonalize $A,B$ separately.

Therefore, unlike the $\mathcal{N}\!=\!2$ proposal for dual ABJM, our $\mathcal{N}\!=\!3$ version does not have an exotic branch in the moduli space in which $M=0$ and $\phi_1,\phi_2$ do not commute \cite{RodriguezGomez:2009ae}. The crucial difference is (\ref{super-2}): without $[\phi_1,\phi_2]$ inside the commutator, $M=0$ will trivialize this condition while not in our case.

\section{The index for the dual ABJM}

The dual ABJM model has four global $U(1)$ symmetries, which we parametrize as $h_1,h_2,h_3,h_4$.
Our notation is such that $h_3,h_4$ form the Cartans of $SO(4)$
R-symmetry, had there been such an enhancement somehow.\footnote{They are denoted by $h_1,h_2$ in \cite{Choi:2008za}.} Our BPS relation is $\epsilon=h_3+j_3$, and the global $U(1)_b$ charge whose associated gauge non-invariance has to be screened by monopole operators is $\frac{h_1+h_2}{2}$.
\begin{table}[t]\label{charges}
$$
\begin{array}{c|c|cc|cc|cc}
  \hline{\rm fields}&U(N)\times U(N)&h_1&h_2&h_3& h_4&j_3&\epsilon\\
  \hline A&(N,\bar{N})&\frac{1}{2}&\frac{1}{2}&-\frac{1}{2}&\frac{1}{2}&0&\frac{1}{2}\\
  B&(\bar{N},N)&-\frac{1}{2}&-\frac{1}{2}&-\frac{1}{2}&\frac{1}{2}&0&\frac{1}{2}\\
  \psi_{\pm}&(N,\bar{N})&\frac{1}{2}&\frac{1}{2}&\frac{1}{2}&\frac{1}{2}&\pm\frac{1}{2}&1\\
  \chi_{\pm}&(\bar{N},N)&-\frac{1}{2}&-\frac{1}{2}&\frac{1}{2}&\frac{1}{2}&\pm\frac{1}{2}&1\\
  \hline\phi_1&({\rm adj},1)&\frac{1}{2}&-\frac{1}{2}&-\frac{1}{2}&-\frac{1}{2}&0&\frac{1}{2}\\
  \phi_2&({\rm adj},1)&-\frac{1}{2}&\frac{1}{2}&-\frac{1}{2}&-\frac{1}{2}&0&\frac{1}{2}\\
  \Psi_{1\pm}&({\rm adj},1)&\frac{1}{2}&-\frac{1}{2}&\frac{1}{2}&-\frac{1}{2}&\pm\frac{1}{2}&1\\
  \Psi_{2\pm}&({\rm adj},1)&-\frac{1}{2}&\frac{1}{2}&\frac{1}{2}&-\frac{1}{2}&\pm\frac{1}{2}&1\\
  \hline A_\mu,\tilde{A}_\mu&&0&0&0&0&(1,0,-1)&1\\
  \hline
\end{array}
$$
\caption{charges of fields in dual ABJM}
\end{table}
The BPS fields in the free field theory are $\bar{A}$, $\bar{B}$, $\bar\phi_i$,
$\psi_{+}$, $\chi_{+}$, $\Psi_{i+}$ and derivatives $D_{++}$. The index over
modes (or letters) is defined as
\begin{equation}
 f_R(x,y_1,y_2)={\rm tr}\left[(-1)^Fe^{-\beta^\prime \{Q,S\}}x^{\epsilon+j_3}y_1^{\frac{h_1-h_2}{2}}y_2^{h_4}\right]\ ,
\end{equation}
where the trace is taken over the normal modes (not over the full space of local gauge invariant operators.)
The letter indices in bi-fundamental, anti-bi-fundamental and two
adjoint representations in the free theory are given by
\begin{equation}
  f_+=f_-=\frac{x^{\frac{1}{2}}y_2^{-\frac{1}{2}}-x^{\frac{3}{2}}y_2^{\frac{1}{2}}}{1-x^2}
  \equiv f\ ,\ \ g=\frac{x^{\frac{1}{2}}y_2^{\frac{1}{2}}
  -x^{\frac{3}{2}}y_2^{-\frac{1}{2}}}{1-x^2}\left(y_1^{\frac{1}{2}}+y_1^{-\frac{1}{2}}\right)\ ,\ \
  \tilde{g}=0\ ,
\end{equation}
respectively.

Monopole operators in $U(N)\times U(N)$ come with fluxes
$H=\{n_1\geq n_2\geq\cdots n_N\}$,
$\tilde{H}=\{\tilde{n}_1\geq\tilde{n}_2\geq\cdots\geq\tilde{n}_N\}$
given by 2 sets of non-increasing integers. The way we calculate
the index is by localization, closely following the analogous
calculation for the ABJM theory \cite{Kim:2009wb}. See also
\cite{Kim:2009ia,Imamura:2009hc} for studies on the index with
monopoles. In particular, we work in the path integral
representation of the index, which refers to the Euclidean QFT on
$S^2\times S^1$. We can deform the action of this theory by
introducing the chiral part of the vector multiplet fields in
$U(N)\times U(N)$, which is exact with our supercharge $Q$ and has
dimension $3$ to preserve the scale invariance. The chiral part of
the superpotential is also $Q$ exact and can be turned off,
leaving the anti-chiral part of the superpotential only in the
action. The resulting path integral can be calculated exactly in
the limit in which the former deformation dominates over the
remaining parts of the action. One can easily check that the
remaining anti-chiral part of the superpotential does not affect
the calculation.\footnote{Since the calculation appears
insensitive to the detailed form of the superpotential, one may
wonder if same calculations can be done in the $\mathcal{N}\!=\!2$
theory with either (\ref{super-original}) or zero superpotential.
It is not clear if (\ref{super-original}) is at a superconformal
fixed point, or if it can flow to a fixed point of the type
studied in \cite{Chang:2010sg}. If it does, and if that point is
continuously connected with our $\mathcal{N}\!=\!3$ fixed point,
the index from the former theory would be the same as ours. If it
is not connected with $\mathcal{N}\!=\!3$ theory, we do not expect
that the index computation followed in the main text applies for
$k=1$ case. This is because there are in general nontrivial
quantum corrections in $\mathcal{N}\!=\!2$ case, especially for
D-terms whose structure is crucial for the index computation and
we do not have a proper understanding of these. This is in
contrast with $\mathcal{N}\!=\!3$ where we have good control in
D-terms and F-terms. The case with zero superpotential corresponds
to an unstable fixed point. The fields here acquire anomalous
dimensions \cite{Gaiotto:2007qi} so that the $Q$ exact deformation
introduced in \cite{Kim:2009wb} could break dilatation symmetry.}
To explain the final result, it is convenient to introduce the
following mode indices
\begin{eqnarray}
  &&f_m=\sum_{i,j=1}^Nx^{|n_i\!-\!\tilde{n}_j|}f\left(e^{-i(\alpha_i\!-\!\tilde\alpha_j)}+
  e^{i(\alpha_i\!-\!\tilde\alpha_j)}\right)\\
  &&f_{adj}=\sum_{i,j=1}^Nx^{|n_i\!-\!n_j|}\left[g-
  (1-\delta_{n_in_j})\right]e^{-i(\alpha_i\!-\!\alpha_j)}\ ,\ \
  \tilde{f}_{adj}=-\sum_{i,j=1}^Nx^{|\tilde{n}_i\!-\!\tilde{n}_j|}(1-\delta_{\tilde{n}_i\tilde{n}_j})
  e^{-i(\tilde\alpha_i\!-\!\tilde\alpha_j)},\nonumber
\end{eqnarray}
where $f_m$ is the contribution from bi-fundamental and anti-bifundamentals, $f_{adj}$, $\tilde{f}_{adj}$ that from the two adjoint matters and modes in the vector multiplets.
The full index with given monopole charge $H$, $\tilde {H}$ is given by
\begin{eqnarray}
  I_{H,\tilde{H}}&=&\frac{1}{({\rm symmetry})}\
  x^{\frac{1}{2}\sum_{i,j=1}^N|n_i\!-\!\tilde{n}_j|-\sum_{i<j}|\tilde{n}_i\!-\!\tilde{n}_j|}\\
  &&\int_0^{2\pi}\prod_{i=1}^N\left[\frac{d\alpha_id\tilde\alpha_i}{(2\pi)^2}\right]
  \prod_{i<j:\ n_i\!=\!n_j}\left[2\sin\frac{\alpha_i\!-\!\alpha_j}{2}\right]^2
  \prod_{i<j:\ \tilde{n}_i\!=\!\tilde{n}_j}\left[2\sin\frac{\tilde\alpha_i\!-\!\tilde\alpha_j}{2}\right]^2
  e^{ik\sum_{i=1}^N(n_i\alpha_i-\tilde{n}_i\tilde\alpha_i)}\nonumber\\
  &&\exp\left[\sum_{n=1}^\infty\frac{1}{n}\left(f_m(x^n,y_1^n,y_2^n,n\alpha,n\tilde\alpha)
  +f_{adj}(x^n,n\alpha,n\tilde\alpha)+\tilde{f}_{adj}(x^n,n\alpha,n\tilde\alpha)\right)\right]\ .\nonumber
\end{eqnarray}
Se also \cite{Kim:2009wb} for detailed explanations on similar calculations. An important difference with the ABJM, apart from the different expressions for $f_m,f_{ad},\tilde{f}_{adj}$, is the zero point energy on the first line. The first term comes from the bi-fundamental matters, and compared to ABJM there is a factor of $\frac{1}{2}$ since the number of chiral multiplets is reduced from 4 to 2. The second term comes
from the vector multiplet in the second $U(N)$, which is the same as ABJM. In ABJM, there is another term $-\sum_{i<j}|n_i\!-\!n_j|$ coming from vector multiplet in first $U(N)$. In the dual ABJM, the
two adjoint matters provide exactly the opposite contribution to cancel this.

In the large $N$ limit keeping energy finite, we do the large $N$ integration over distributions for holonomies $\alpha_i,\tilde\alpha_i$ which do not support magnetic flux. Just like the case studied in \cite{Kim:2009wb}, the index then factorizes as
\begin{equation}
  I_{free}I_>I_<\ ,
\end{equation}
as we explain now. The first part is the result of Gaussian
integration,
\begin{equation}\label{free}
  I_{free}=\prod_{n=1}^\infty\frac{1}{1-g-f^2}
  =\prod_{n=1}^\infty\frac{(1-x^{2n})^2}{(1-x^ny_2^{-n})\left(1-(xy_2)^{\frac{1}{2}}
  (y_1^{\frac{1}{2}}+y_1^{-\frac{1}{2}})(1-x^2)-x^3y_2\right)}
\end{equation}
and is exactly the part calculated in \cite{Choi:2008za} without monopole operators. $I_>$, $I_<$, apart from the zero point energy part that we explain later, are given as follows. In terms of
\begin{eqnarray}
  {\rm\bf f}_{ij}^m&=&\left(x^{|n_i\!-\!\tilde{n}_j|}-
  x^{|n_i|\!+\!|\tilde{n}_j|}\right)f\left(e^{-i(\alpha_i\!-\!\tilde\alpha_j)}
  +e^{i(\alpha_i\!-\!\tilde\alpha_j)}\right)\nonumber\\
  {\rm\bf f}^{adj}_{ij}&=&\left[(g-1+\delta_{n_in_j})x^{|n_i\!-\!n_j|}-(g-1)x^{|n_i|\!+\!|n_j|}\right]
  e^{-i(\alpha_i\!-\!\alpha_j)}\\
  \tilde{\rm\bf f}^{adj}_{ij}&=&-\left[(1-\delta_{\tilde{n}_i\tilde{n}_j})x^{|\tilde{n}_i\!-\!\tilde{n}_j|}
  -x^{|\tilde{n}_i|\!+\!|\tilde{n}_j|}\right]e^{-i(\tilde\alpha_i\!-\!\tilde\alpha_j)}\ ,\nonumber
\end{eqnarray}
$I_>$ is given by
\begin{eqnarray}
  &&\frac{1}{({\rm symmetry})}\int_0^{2\pi}\prod_i\left[\frac{d\alpha_i}{2\pi}\right]\prod_j\left[\frac{d\tilde\alpha_j}{2\pi}\right]
  \prod_{i<j:\ n_i\!=\!n_j}\left[2\sin\frac{\alpha_i\!-\!\alpha_j}{2}\right]^2
  \prod_{i<j:\ \tilde{n}_i\!=\!\tilde{n}_j}\left[2\sin\frac{\tilde\alpha_i\!-\!\tilde\alpha_j}{2}\right]^2
  e^{ik\sum_i(n_i\alpha_i-\tilde{n}_i\tilde\alpha_i)}\nonumber\\
  &&\exp\left[\sum_{n=1}^\infty\frac{1}{n}\left(\sum_{i,j}{\rm\bf f}_{ij}^m(x^n,y_1^n,y_2^n,e^{in\alpha},e^{in\tilde\alpha})+\sum_{i,j}{\rm\bf f}^{adj}_{ij}
  (x^n,e^{in\alpha})+\sum_{i,j}\tilde{\rm\bf f}^{adj}_{ij}(x^n,e^{in\tilde\alpha})\right)\right]
\end{eqnarray}
where $n_i,\tilde{n}_j$ are the positive part of the fluxes in $H,\tilde{H}$, and
$\alpha,\tilde\alpha$ are associated holonomy. $I_<$ is given by a similar expression using negative fluxes in $H,\tilde{H}$ only. The whole integrand factorizes into $I_>$ and $I_<$ as shown above, as ABJM. The only
remaining thing that one has to show is whether the zero point energy
\begin{equation}
  x^{\frac{1}{2}\sum_{i,j=1}^N|n_i\!-\!\tilde{n}_j|-\sum_{i<j}|\tilde{n}_i\!-\!\tilde{n}_j|}
\end{equation}
factorizes to two parts, where each part refers to positive or
negative fluxes only.
\begin{equation}
  \epsilon_0=\frac{1}{2}\sum_{i,j=1}^N|n_i\!-\!\tilde{n}_j|-
  \sum_{i<j}|\tilde{n}_i\!-\!\tilde{n}_j|\ .
\end{equation}
We now explain this factorization. Obviously there are many pairs
which connect two positive or two negative fluxes, which factorize.
We only have to consider the pairs connecting one positive and one
negative flux. Its factorization is almost clear since
$|x-y|\!=\!|x|\!+\!|y|$ if $x\!>\!0\!>\!y$. The only reason why we
have to be careful is that the positive flux part may acquire
constant coefficient depending on summation over $U(1)$'s with
negative flux. To study this, let us define $N_\pm$ and
$\tilde{N}_\pm$ to be the number of $U(1)$ Cartans which support
positive/negative fluxes in two gauge groups. Then the contribution
from pairs connecting positive and negative fluxes is
\begin{eqnarray}
  &&\frac{1}{2}\left((N\!-\!\tilde{N}_+)\sum|n_i^+|+(N\!-\!N_-)
  \sum|\tilde{n}_i^-|+(N\!-\!\tilde{N}_-)\sum|n_i^-|+
  (N\!-\!N_+)\sum|\tilde{n}_i^+|\right)\nonumber\\
  &&+(N\!-\!\tilde{N}_+)\sum|\tilde{n}_i^+|+
  (N\!-\!\tilde{N}_-)\sum|\tilde{n}_i^-|\ ,
\end{eqnarray}
where we included contributions from pairs with one zero flux and
one nonzero flux. From this expression, the coefficients of the
summations involving positive/negative fluxes only refer to the
number of $U(1)$'s with positive/negative fluxes, respectively. This
proves the factorization of zero point energy, and they should be
included in the definition of $I_>$ and $I_<$. We also multiply $ y_3^{\frac{h_1+h_2}{2}}\sim y_3^{\frac{k}{2}\sum_i n_i}$ to weight the operators with their $\frac{h_1+h_2}{2}$ charges.

Now we consider the case with $k\!=\!1$, which has the moduli space $\mathbb{C}^4$ and is thus suspected to be the dual description of ABJM. We can either compare the above index with that for the ABJM theory at $k\!=\!1$, or that over supersymmetric gravitons in $AdS_4\times S^7$ (which were checked to be the same \cite{Kim:2009wb}). Since there is a complete factorization $I_{free}I_>I_<$, it is straightforward to show that $I_{free}$, $I_>$ and $I_<$ should agree with indices from gravitons with zero KK-momenta, positive momenta, negative momenta separately \cite{Kim:2009wb}. Note also that, for $k\!=\!1$, the M-theory geometry is completely smooth so that the above supergravity index suffices. However, it has been shown in \cite{Choi:2008za} that $I_{free}$ does not agree with that from gravitons with zero KK-momenta. So whatever
happens to $I_>$ and $I_<$ in the monopole sector, the two indices cannot agree.

In \cite{RodriguezGomez:2009ae}, the disagreement of the index in \cite{Choi:2008za} was suspected to be due to the presence of an exotic branch in the moduli space if one analyzes the classical moduli space with (\ref{super-original}). In the previous section, we re-analyzed the moduli space with the superpotential
(\ref{modified-ABJM}) at the $\mathcal{N}\!=\!3$ fixed point and
showed the absence of this exotic branch. Therefore, the mismatch
between the two indices should be somehow explained
differently.\footnote{Of course one explanation would simply be
regarding this as the evidence that dual ABJM at $k\!=\!1$ does not
describe M2-branes in $\mathbb{R}^8$.}

At this point, we explain a feature of the index $I_{free}$ which
does not seem to be emphasized in \cite{Choi:2008za}. The ranges of
the chemical potentials $x,y_1,y_2$ should be such that the trace
over the space of modes converges. This is guaranteed by taking $x$
to be sufficiently small. Investigating the charges of all fields,
one obtains the following conditions
\begin{equation}\label{chemical-allowed}
  xy_2^{\pm 1}<1\ ,\ \ x^3y_2^{\pm 1}<1\ ,\ \ xy_1^{\pm 1}y_2^{\pm 1}
  <1\ ,\ \ x^3y_1^{\pm 1}y_2^{\pm 1}<1\ .
\end{equation}
The equations involving $x^3$, coming from fermionic letters, are
automatically implied once we accept the other conditions from
bosons. With this understanding, let us review the calculation of
large $N$ saddle point calculation which led us to $I_{free}$ in
(\ref{free}). The basic idea is that the integral over $N$ holonomies
reduces to a Gaussian integration over the distribution of them. To obtain (\ref{free}),
all the eigenvalues of the matrix appearing in the quadrature in the
exponent should be positive. This is guaranteed for a sufficiently
small $x$ (which is analogous to the low temperature limit of the
partition function). We should also check if some eigenvalues of
this matrix can turn to be negative for some values of chemical
potentials. Since (\ref{free}) is nothing but the inverse of the
determinant of this matrix, the sign change of any eigenvalue can be
detected as the divergence of $I_{free}$. So let us have a look at
the denominator of (\ref{free}). The first factors involving
$(1-(xy_2^{-1})^n)$ never approaches zero in the allowed range
(\ref{chemical-allowed}). However, the second factor
\begin{equation}
  \left(1-(xy_2)^{\frac{1}{2}}(y_1^{\frac{1}{2}}+y_1^{-\frac{1}{2}})(1-x^2)
  -x^3y_2\right)
\end{equation}
can approach zero in the allowed range (\ref{chemical-allowed}), as
follows. Firstly, it is much simpler to consider the simple case
with $y_1\!=\!y_2\!=\!1$, in which case the allowed region is simply
$x<1$. Then the above factor becomes
\begin{equation}
  1-2x^{\frac{1}{2}}+2x^{\frac{5}{2}}-x^3=
  (1-x)\left(1-2x^{\frac{1}{2}}+x-2x^{\frac{3}{2}}+x^2\right)\ .
\end{equation}
The expression appearing in the second parenthesis monotonically
decreases from $1$ to $-1$ in the range $0\!<\!x\!<\!1$, which
becomes zero approximately at $x^{\frac{1}{2}}\approx 0.5310$.
Beyond this value, one eigenvalue becomes negative and the previous
saddle point at which the distribution function is uniform seizes to
be a minima. Instead one should find a new saddle point at which the
distribution function is not uniform \cite{Aharony:2003sx}.
The index undergoes a transition to a `deconfined' phase.

At the above deconfined phase, the (index version of) free energy is proportional
to $N^2$ \cite{Aharony:2003sx}, which is much larger than $N^{\frac{3}{2}}$ which
we expect in the high temperature phase of M2 branes. We should emphasize that,
in all Chern-Simons-matter type gauge theory models for M2 branes, the $N^2$ weakly
coupled degrees of freedom should not all appear in the strongly interacting regime
(say at $k\!=\!1$) for these theories to describe M2 branes. Actually the index never
deconfines in the case of ABJM \cite{Kim:2009wb}. The fact that $N^2$ degrees of freedom
appear in the index for the `dual ABJM' seems to be in sharp contrast with the M2 brane picture.

\section*{Acknowledgments}

The work of S.K. is supported in part by the Research Settlement Fund
for the new faculty of Seoul National University.
The work of J.P. is supported in part by KOSEF Grant
R01-2008-000-20370-0, by the National Research Foundation of
Korea(NRF) grant funded by the Korea government(MEST) with  the
grant number  2009-0085995 and by the grant number 2005-0049409
through the Center for Quantum Spacetime(CQUeST) of Sogang
University.

\end{document}